# Incorporation of Magnesium into GaN Regulated by Intentionally Large Amounts of Hydrogen during Growth by MOCVD

*Anelia Kakanakova-Georgieva\*, Alexis Papamichail, Vallery Stanishev and Vanya Darakchieva*

Anelia Kakanakova-Georgieva, Alexis Papamichail, Vallery Stanishev, Vanya Darakchieva
Division of Semiconductor Materials, Department of Physics, Chemistry and Biology (IFM),
Linköping University, SE-58183 Linköping, Sweden;
Center for III-Nitride Technology, C3NiT – Janzén,
Linköping University, SE-58183 Linköping, Sweden
E-mail: anelia.kakanakova@liu.se

Vallery Stanishev, Vanya Darakchieva
Terahertz Materials Analysis Center, THeMAC,
Linköping University, SE-58183 Linköping, Sweden

Vanya Darakchieva
Solid State Physics and NanoLund,
Lund University, P. O. Box 118, 221 00 Lund, Sweden

Metalorganic chemical vapor deposition (MOCVD) of GaN layers doped with Mg atoms to the recognized optimum level of *[Mg] ~ $2\times10^{19}$ cm$^{-3}$* has been performed. In a sequence of MOCVD runs, operational conditions, including temperature and flow rate of precursors, have been maintained except for intentionally larger flows of hydrogen carrier gas fed into the reactor. By employing the largest hydrogen flow of 25 slm in this study, the performance of the *as-grown* Mg-doped GaN layers has been certified by a room-temperature hole concentration of *p ~ $2\times10^{17}$ cm$^{-3}$* in the absence of any thermal activation treatment. Experimental evidence is delivered that the large amounts of hydrogen during the MOCVD growth can regulate the incorporation of the Mg atoms into GaN in a significant way so that MgH complex can co-exist with a dominant and evidently electrically active isolated $Mg_{Ga}$ acceptor.

## 1. Introduction

Magnesium (Mg) doping of GaN achieves epitaxial layers of *p*-type conductivity for developing GaN-based devices such as light-emitting and vertical *pn* diodes. In more recent years, developing GaN vertical *pn* diodes consolidates expanding research efforts as it can promote power electronics applications which take advantage of superior material properties of wide band gap GaN as compared to Si and SiC.[1-4]

Magnesium has proved as the only effective *p*-type dopant for GaN. Yet, the performance of the *p*-type GaN obeys principal limitations related to the high ionization energy of the Mg acceptor in GaN, passivation of the Mg acceptor enabled by co-incorporated hydrogen in the MOCVD processes and low formation energy of other compensating point defects, particularly nitrogen vacancy ($V_N$). The magnesium acceptor involves substitutional Mg atom on the Ga site in the GaN crystal lattice ($Mg_{Ga}$). A typical Mg atomic concentration in *p*-type GaN by MOCVD, including for the purpose of its integration in power device structures,[1,2] assumes a nominal value of *[Mg] ~ (2-3)×10$^{19}$ cm$^{-3}$*. This Mg atomic concentration correlates to a reduced hole concentration of *p ~ (4-6)×10$^{17}$ cm$^{-3}$*,[1,5] which is a result to the high ionization energy of the Mg acceptor in GaN. Room-temperature Hall-effect measurements of the hole concentration as a function of the Mg atomic concentration in the doping range of ~ *3×10$^{18}$ - 1×10$^{20}$ cm$^{-3}$* often found that the hole concentration peaks at a threshold value of ~ *(2-3)×10$^{19}$ cm$^{-3}$* and decreases for larger values.[4,5] Moreover, a measurable hole concentration normally requires putting in place a post-growth thermal annealing of the Mg-doped GaN in a nitrogen atmosphere. The post-growth thermal annealing causes dissociation of MgH complexes and hydrogen out diffusion which renders the Mg acceptors electrically active. Overall, well-optimized material growth and thermal annealing conditions, as well as contact technology, may improve the performance of the *p*-type GaN layers whereby typical hole mobility and hole conductivity at room temperature can measure around 10 cm$^2$/Vs and less than 1 (Ω cm)$^{-1}$, respectively.[5,6]

We report on *as-grown* GaN layers doped with Mg atoms to the recognized threshold value of *[Mg]~2×10$^{19}$ cm$^{-3}$* which performance can be certified by a room-temperature hole concentration of $p \sim 2\times10^{17}$ $cm^{-3}$ in the absence of any thermal activation treatment. The advantageous performance of the *as-grown* GaN layers is a merit to MOCVD processes operated with intentionally large amounts of hydrogen. We present experimental evidence that the large amounts of hydrogen during the MOCVD growth can regulate the incorporation of the Mg atoms into GaN in a significant way so that MgH complexes can co-exist with a dominant and evidently electrically active isolated $Mg_{Ga}$ acceptors.

Our motivation for operating MOCVD of GaN and Mg doping with intentionally large amounts of hydrogen acknowledges the outcome of first principles calculations about incorporation of dominant defects in *p*-type GaN including Mg acceptor ($Mg_{Ga}$), interstitial hydrogen ($H_i$), and nitrogen vacancy ($V_N$) and examination of the plot of the formation energy of these defects in their charge state of $Mg_{Ga}^-$, $H_i^+$, and $V_N^+$ versus the equilibrium Fermi energy position, as well as their corresponding equilibrium concentrations for both H-rich and H-free limit.[7] It has been found that (i) formation of $H_i^+$ is energetically most favorable and thereby it is the dominant compensating donor defect; (ii) passivation of Mg acceptor by H shifts the Fermi level high in the band gap and thereby causes an increase in the formation energy for other compensating donor defects, particularly $V_N^+$; and (iii) "*compared to the hydrogen-free case, the Mg concentration is increased and the $V_N$ concentration is decreased*".[7] The herein referred phenomena, being directly related to a hydrogen abundance in GaN, rationalize a beneficial role of hydrogen in p-type doping of GaN by magnesium. [7,8] An acknowledged strategy in "*defeating compensation*" in wide band gap semiconductors, such as GaN, relates to engaging hydrogen as a provisional passivating agent during growth which is subsequently to be removed.[9] Conditions under which hydrogen can be used to control doping in GaN, and in semiconductors in general, has largely been discussed from the perspective of first-principles studies. Here, we give an experimental aspect to the perceived beneficial role of hydrogen in the p-type doping of GaN by magnesium.

## 2. Results and Discussion

The Mg doping level in the *as-grown* GaN layers is determined by the $Cp_2Mg$/TMGa precursor ratio, herein covering the range of low $10^{18}$ to high $10^{19}$ $cm^{-3}$ (**Figure 1a**.) These layers have been grown with a same hydrogen flow of 19 slm. An *as-grown* GaN layer with *[Mg] ~ 1.6×10$^{19}$ cm$^{-3}$* (H19 in Fig. 1a) has been assigned as a reference and further represented by its surface morphology of low root-mean-square (rms) roughness of about 0.15 nm for a scan area of 5x5 μm² (**Figure 1b**) and SIMS depth profiles of incorporated Mg, H, Si, O, and C atoms (**Figure 1c**). Besides confirming the co-incorporation of H at about the same atomic concentration as that of Mg in the *as-grown* GaN layer, incorporation of residual silicon and oxygen impurity atoms at the SIMS detection limit, and residual carbon atoms at the low level of ~ $1 \times 10^{16}$ cm$^{-3}$ has also been confirmed. We point out that the atomic concentration of Mg in the reference GaN layer H19 is below the threshold value beyond which the atomic concentrations of Mg and H decouple (Figure 1a) and a saturation in the incorporation of MgH complex becomes relevant upon high levels of Mg doping.[10, 11]

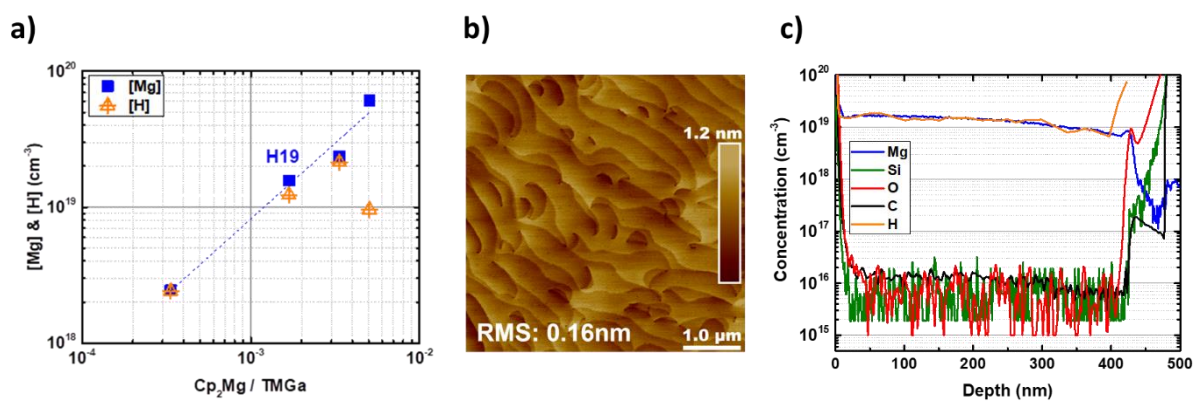

**Figure 1.** a) Atomic concentration of Mg and H in *as-grown* GaN layers by SIMS vs. Cp$_2$Mg/TMGa precursors ratio under hydrogen flow of 19 slm, b) AFM image 5×5μm² representative of the surface morphology of the GaN layer from the reference sample H19, c) SIMS depth profiles of Mg, H, Si, O, and C atomic concentration in the sample H19. Due to transient effects, concentrations gradients within 40 nm below the surface are irrelevant. The SIMS background levels in GaN correspond to [Mg] ~ $2 \times 10^{16}$ cm$^{-3}$, [H] ~ $\times 10^{17}$ cm$^{-3}$, [Si] ~ $6 \times 10^{15}$ cm$^{-3}$, [C] ~ $2 \times 10^{15}$ cm$^{-3}$, and [O] ~ $6 \times 10^{15}$ cm$^{-3}$. The GaN layer has been grown directly onto an AlN nucleation layer. The SIMS concentration levels in the AlN and SiC substrate are not accurate.

Next, the reference H19 has been repeated by maintaining the same Cp$_2$Mg/TMGa precursor ratio, except for the hydrogen flow being set to ever larger amounts of 21, 23, and 25 slm, and the layers have been accordingly labeled as H21, H23, and H25. The individual Mg depth profiles measured in each of the layers have been plotted in **Figure 2a**. A clear up-shift of the Mg concentration profiles for the group of layers H21-H25 compared to H19 may be noticed and interpreted as originating in the exposure of GaN to larger amounts of hydrogen during the MOCVD growth.

Before continuing a further discussion, we note that we operate a horizontal-type MOCVD reactor. Increasing the amount of hydrogen fed into the reactor may cause a shift of the deposition profile downstream along with a dilution of the precursors and a reduction of the growth rate at a same substrate location which may impact the incorporation of Mg occupying Ga sites. For that reason, we have imposed a limit on the amount of hydrogen fed into the reactor. Out of total of 50 slm available, the $H_2$ flow has only been increased as of 19-> 21-> 23-> 25 slm so that to render adequate experiments. It can be reasoned by comparing the thickness of the GaN layers which have been obtained in a same deposition time. The thickness of each of the GaN layers has been extracted from its SIMS depth profile according to the Ga marker signal and the thickness value reads ~ 438-> 460-> 423-> 417 nm, respectively. The GaN thickness variation, defined by the expression: (standard deviation)/(mean value)x100 [%], is down to about 4% for this set of runs.

Hydrogen-rich condition are expected to enhance passivation of Mg atoms incorporated into GaN. It is considered as a beneficial effect by shifting the Fermi level high in the band gap thereby suppressing compensating point defects.[7] At this point, an increase in the Fermi energy decreases the formation energy of acceptors ($Mg_{Ga}$) and hence increases their concentration (see Figure 1a in ref.no.7 for details about formation energy of $Mg_{Ga}^-$ and $H_i^+$, and hence their concentrations). Despite the overall joint co-incorporation of Mg and H, a certain decoupling in the Mg and H atomic concentration in each of the individual *as-grown* GaN layer occurs. It has been exemplified by plotting the SIMS profiles of the Mg and H atomic concentrations in the *as-grown* GaN layer H25 (**Figure 2b**). On average, the distribution of the Mg atoms throughout this GaN layer is at a concentration of ~$1.8\times10^{19}$ cm$^{-3}$ and that of H at ~$1.4\times10^{19}$ cm$^{-3}$ resulting in a difference between the incorporated Mg and H atoms of ~$4\times10^{18}$ cm$^{-3}$.

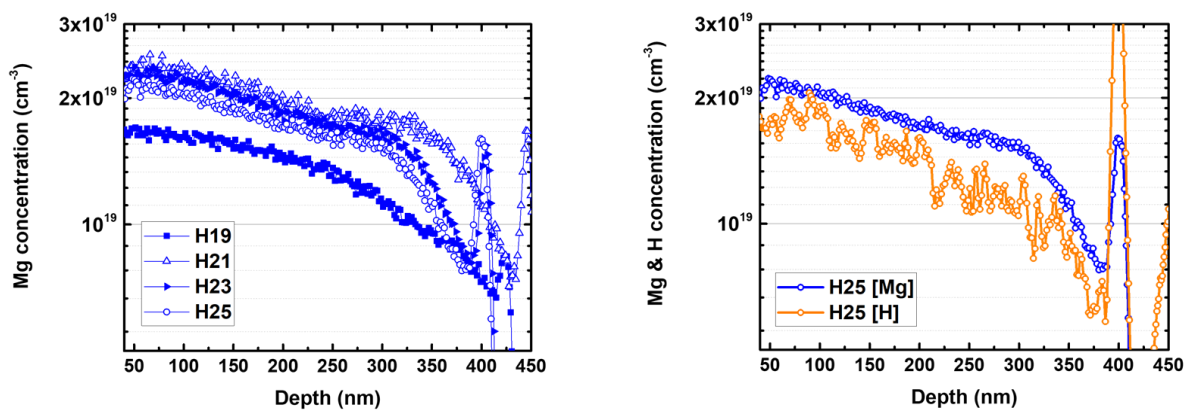

**Figure 2.** a) atomic concentration of Mg by SIMS in *as-grown* GaN layers H19, H21, H23 and H25, b) the Mg atomic concentration in the *as-grown* GaN layer H25 is plotted together with the H atomic concentration.

For the plot in **Figure 3**, the relative difference between the incorporated Mg and H atoms, ([Mg]-[H])/[Mg], in each of the *as-grown* GaN layers has been assigned as a parameter. The value of this parameter increases by a factor of 5 as larger amounts of hydrogen have been employed for the Mg doping of GaN. Consequently, our data is consistent with the theoretically predicted expectation for the Mg incorporation being largely affected by abundance of hydrogen atoms.[7] The abundance of hydrogen atoms under the operational conditions in our MOCVD experiments - which in all maintain same flow rate of precursors and temperature - apparently relate to the introduction of larger amounts of hydrogen and up to 25 slm. This large amount of hydrogen proves pertinent in giving rise to molecular hydrogen dissociation at the growing surface and production of atomic hydrogen at a more significant rate.[12] This argument refers to a previous study concluding "*hydrogenation*" of AlN grown in the same MOCVD reactor under a set of conditions involving same large amount of hydrogen.[12]

Noteworthy, the hole concentration in the *as-grown* "*hydrogenated*" GaN layers scales-up linearly with the relative difference between the incorporated Mg and H atoms and in a significant way over an order of magnitude, from ~ $7.6 \times 10^{15}$ to $8.6 \times 10^{16}$ cm$^{-3}$ (**Figure 3**). This result is thus suggestive for incorporation of evidently non-passivated isolated Mg$_{Ga}$ acceptor, co-existing with the dominant MgH complex, which can be electrically active in *as-grown* GaN. The measured hole concentration expressed as a fraction of these non-passivated Mg acceptors, i.e., ~ $7.6\times10^{15}$ cm$^{-3}$/$6\times10^{17}$ cm$^{-3}$ (H19) and ~ $8.6\times10^{16}$ cm$^{-3}$/$4\times10^{18}$ cm$^{-3}$ (H25), respectively, gives an estimation of a few percent and therefore corresponds to the expected doping efficiency of Mg acceptor in GaN, which is below 2%.

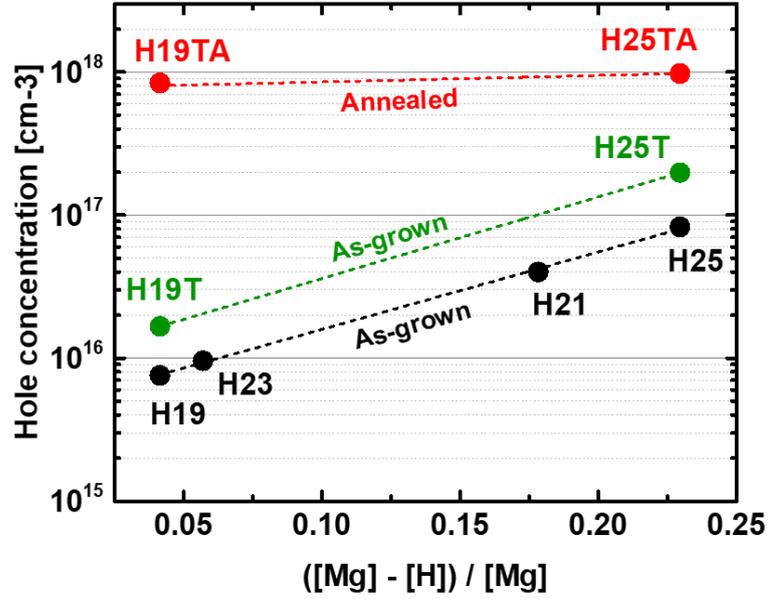

**Figure 3.** Hole concentration measured in layers H19 - H25 (*as-grown* GaN:Mg on AlN/SiC template), H19T & H25T (*as-grown* GaN:Mg on GaN/AlN/SiC template) and H19TA & H25TA (annealed GaN:Mg on GaN/AlN/SiC template).

We exclude that the measured hole concentration in the *as-grown* GaN might be a result of any *in-situ* annealing in effect during the cooling down procedure. The cooling down procedure has been performed under hydrogen rich conditions. In each of the GaN growth runs, same H$_2$ flow of 19 slm continued during the entire cooling down stage of the overall MOCVD process.

To complete the study, we note that, in principle, a performance enhancement can be achieved by optimizing the structural properties of the Mg-doped GaN layer by its deposition on GaN/AlN/SiC template instead of AlN/SiC template (the layers have been accordingly labeled as H19T and H25T), and by further post-growth annealing (the layers have been accordingly labeled as H19TA and H25TA). Comparative hole concentration values have been plotted in **Figure 3** with other transport property value of hole mobility and resistivity summarized in **Table 1.**

**Table 1.** Summary of transport properties for Mg-doped GaN layers labeled in correspondence to the plot in Figure 3.

|  | *as-grown* GaN:Mg/AlN/SiC | *as-grown* GaN:Mg/GaN/AlN/SiC | *annealed* GaN:Mg/GaN/AlN/SiC |
|---|---|---|---|
|  | p (cm$^{-3}$) / ρ (Ω.cm) / mobility (cm$^{-2}$/V s) | p (cm$^{-3}$) / ρ (Ω.cm) / mobility (cm$^{-2}$/V s) | p (cm$^{-3}$) / ρ (Ω.cm) / mobility (cm$^{-2}$/V s) |
| H$_2$ = 19 slm | 7.6×10$^{15}$ / 92 / 9.17 (H19) | 1.7×10$^{16}$ / 42 / 8.95 (H19T) | 8.4×10$^{17}$ / 0.8 / 9.69 (H19TA) |
| H$_2$ = 25 slm | 8.6×10$^{16}$ / 11.2 / 6.50 (H25) | 2.0×10$^{17}$ / 3.7 / 8.51 (H25T) | 9.8×10$^{17}$ / 0.7 / 8.74 (H25TA) |

Particularly, the outcome from the post-growth annealing can be indicative for revealing a certain framework of incorporation of the Mg acceptors into GaN layers depending on the amount of hydrogen employed in the MOCVD processes in this study. Obviously, the post-growth annealing of the Mg-doped GaN layer grown under the hydrogen flow of 19 slm causes a large increase in the hole concentration which is by a factor of about 50 (**Figure 3**, H19T -> H19TA). The data is therefore consistent with a perception for a large amount of MgH complexes being present in this layer and the incorporated Mg acceptors becoming electrically active only after thermal activation. This effect is less important in the Mg-doped GaN layer grown under the largest amount of hydrogen, and whereas the hole concentration has been marked by an increase, it is by a factor of about 5 (**Figure 3**, H25T -> H25TA).

Overall, same nominal Mg doping level in the GaN layers achieves comparable hole concentration with post-growth annealing. Importantly, however, a certain proportion between evidently non-passivated isolated Mg$_{Ga}$ acceptors and MgH complexes can be established by intentionally increasing the amount of hydrogen during MOCVD in *as-grown*

Mg-doped GaN. Our study therefore implies a pragmatic growth approach that results in achieving a hole concentration value in the range of $p \sim 2\times10^{17}$ $cm^{-3}$ in *as-grown* GaN layers (i.e., H25T) in the absence of any thermal activation treatment. Mg-doped GaN layers with a reduced hole concentration to $p \sim 3\times10^{17}$ $cm^{-3}$ have been integrated into recently reported device structures supporting a 4.9 kV breakdown voltage vertical GaN p–n junction diode[13] and a vertical 3D gallium nitride field-effect transistor.[14]

In summary, we have demonstrated a deliberate involvement of hydrogen in the control of Mg incorporation by an adequate experiment under regular MOCVD conditions in the range of moderate Mg doping and serving pragmatic purposes. The outcome of this study can motivate further inquiries into incorporation of ever larger amounts of atomic hydrogen into GaN.

## 4. Experimental Section

The MOCVD of Mg-doped GaN has been performed on SiC substrates in a horizontal-type hot-wall MOCVD reactor (GR508GFR Aixtron). Different aspects of previous developments in the same reactor related to, e.g., Mg-doped $Al_{0.85}Ga_{0.25}N$ layers with low resistivity at room temperature,[15] strain and morphology compliance during intentional doping of high-Al-content AlGaN layers,[16] and exciton luminescence in AlN triggered by hydrogen and thermal annealing,[12] can be found elsewhere. For this study, trimethylgallium, $(CH_3)_3Ga$ or TMGa, ammonia, $NH_3$, and bis(cyclopentadienyl)magnesium, $Cp_2Mg$, have been implemented as precursors and delivered in a mixture of hydrogen ($H_2$) and nitrogen ($N_2$) gases. The flow rate of the TMGa precursor has been set to 2.21 sccm. The flow rate of the $Cp_2Mg$ precursor, which delivers the Mg dopant atoms, has been varied and, accordingly, the specific values of the $Cp_2Mg$/TMGa precursor ratio can be found in the text. The flow rate of $NH_3$ and $N_2$ has been set to a constant value of 2 slm and 9 slm, respectively. The reactor has been operated at a temperature of 1120 ºC and a pressure of 100 mbar for the growth of GaN. For this study, and in several separate runs, the flow rate of $H_2$ has been set to values of 19, 21, 23, and 25 slm as to achieve a larger exposure of GaN to hydrogen during growth. In each of these runs, same $H_2$ flow of 19 slm continued during the entire cooling down stage of the overall MOCVD process. The surface morphology of the layers has been obtained by atomic force microscopy (AFM) and 5×5μm$^2$ images have been acquired in a tapping mode using Veeco Dimension 3100 Scanning Probe Microscope. Secondary ion mass spectrometry (SIMS) has been employed to obtain the thickness of the GaN layers and to determine the Mg and H atomic concentrations alongside Si, O and C.[17] Hole concentration, mobility and resistivity have been determined by measurements in Van der Paw configuration with a Linseis HCS 1 instrument. Ni/Au (5 nm/ 250 nm) contacts have been deposited by thermal evaporation and annealed at 450 °C in air to render ohmic behavior. The activation of the Mg acceptors was performed by a rapid thermal annealing (RTA) at 900˚C for 10 min under a $N_2$ environment.

**Acknowledgements**

This work is performed within the framework of the competence center for III-Nitride technology, C3NiT - Janzén supported by the Swedish Governmental Agency for Innovation Systems (VINNOVA) under the Competence Center Program Grant No. 2016-05190, Linköping University, Chalmers University of technology, Ericsson, Epiluvac, FMV, Gotmic, Hexagem, Hitachi Energy, On Semiconductor, Saab, SweGaN, and UMS. We further acknowledge support from the Swedish Research Council VR under Award No. 2016-00889, Swedish Foundation for Strategic Research under Grants No. RIF14-055 and No. EM16-0024, and the Swedish Government Strategic Research Area in Materials Science on Functional Materials at Linköping University, Faculty Grant SFO Mat LiU No. 2009-00971. We thank Dr. S. P. Le for metal contact preparation, B. Hult and Prof. N. Rorsman for the RTA annealing and Dr. Jr.-Tai Chen for fruitful discussions.

References

[1]     I. C. Kizilyalli, A. P. Edwards, O. Aktas, T. Prunty, D. Bour, *IEEE Trans Electron Devices* **2015**, *62*, 414.

[2]      Y.Zhang, D. Piedra, M. Sun, J. Hennig, A. Dadgar, L. Yu, T. Palacios, *IEEE Electron Device Lett.* **2017**, *38,* 248.

[3]     H. Amano, *et al*, *J. Phys. D: Appl. Phys.* **2018**, *51,* 163001.

[4]     T. Narita, H. Yoshida, K. Tomita, K. Kataoka, H. Sakurai, M. Horita, M. Bockowski, N. Ikarashi, J. Suda, T. Kachi, Y. Tokuda, J. Appl. Phys. **2020**, *128,* 090901.

[5]     U. Kaufmann, P. Schlotter, H. Obloh, K. Köhler, M. Maier, *Phys. Rev. B* **2000**, *62,* 10867.

[6]     P. Kozodoy, H. Xing, S. P. DenBaars, U. K. Mishra, A. Saxler, R. Perrin, S. Elhamri, W. C. Mitchel, *J. Appl. Phys.* **2000**, *87*, 1832.

[7]     J. Neugebauer, C. G. Van de Walle, *Appl. Phys. Lett.* **1996**, *68,* 1829.

[8]     J. L. Lyons, A. Janotti, C. G. Van de Walle, *Phys. Rev. Lett.* **2012**, *108*, 156403.

[9]     J. A. van Vechten, J. D. Zook, R. D. Hornig, and B. Goldenberg, *Jpn. J. Appl. Phys*. **1992**, *31*, 3662.

[10]    A. Papamichail, A. Kakanakova, E. O. Sveinbjörnsson, A. R. Persson, B. Hult, N. Rorsman, V. Stanishev, S. P. Le, P. O. Å. Persson, M. Nawaz, Jr-Tai Chen, P. P. Paskov, V. Darakchieva, *J. Appl. Phys*. **2022**, DOI: 10.1063/5.0089406.

[11]    A. Castiglia, J.-F. Carlin, N. Grandjean, *Appl. Phys. Lett.* **2011**, *98*, 213505.

[12]    M. Feneberg, N. T. Son, A. Kakanakova-Georgieva, *Appl. Phys. Lett.* **2015**, *106*, 242101.

[13]    H. Ohta, N. Asai, F. Horikiri, Y. Narita, T. Yoshida, T. Mishima, *Jpn. J. Appl. Phys*. **2019**, *58*, SCCD03.

[14]    K. Strempel, F. Römer, F. Yu, M. Meneghini, A. Bakin, H. H. Wehmann, B. Witzigmann, A. Waag, *Semicond. Sci. Technol*. **2020**, *36*, 014002.

[15]    A. Kakanakova-Georgieva, D. Nilsson, M. Stattin, U. Forsberg, Å. Haglund, A. Larsson, E. Janzén, *Phys. Status Solidi RRL* **2010**, *4*, 311.

[16]    D. Nilsson, E. Janzén, A. Kakanakova-Georgieva, *Appl. Phys. Lett.* **2014**, *105*, 082106.

[17]    EAG Laboratories, www.eag.com